\documentclass[aip,reprint]{revtex4-2}

\usepackage{graphicx}
\usepackage{dcolumn}
\usepackage{bm}
\usepackage{xcolor}
\usepackage[utf8]{inputenc}
\usepackage[T1]{fontenc}
\usepackage{mathptmx}

\begin{document}

\preprint{AIP/123-QED}

\title{Perspective on witnessing entanglement in hybrid quantum systems}

\author{Yingqiu Mao}
\email{Authors to whom correspondence should be addressed: myingqiu@ustc.edu.cn and johannes@majer.ch}
\affiliation{ 
Hefei National Laboratory for Physical Sciences at the Microscale, University of Science and Technology of China, Hefei 230026, China}
\affiliation{Shanghai Branch, CAS Center for Excellence in Quantum Information and Quantum  Physics, University of Science and Technology of China, Shanghai 201315, China}

\author{Ming Gong}%
\affiliation{ 
Hefei National Laboratory for Physical Sciences at the Microscale, University of Science and Technology of China, Hefei 230026, China}
\affiliation{Shanghai Branch, CAS Center for Excellence in Quantum Information and Quantum  Physics, University of Science and Technology of China, Shanghai 201315, China}

\author{Kae Nemoto}
\affiliation{National Institute of Informatics, 2-1-2 Hitotsubashi, Chiyoda-ku, Tokyo 101-8430, Japan}

\author{William J. Munro}
\affiliation{NTT Basic Research Laboratories and Research Center for Theoretical Quantum Physics, 3-1 Morinosato-Wakemiya, Atsugi, Kanagawa 243-0198, Japan}
\affiliation{National Institute of Informatics, 2-1-2 Hitotsubashi, Chiyoda-ku, Tokyo 101-8430, Japan}

\author{Johannes Majer}
\affiliation{ 
Hefei National Laboratory for Physical Sciences at the Microscale, University of Science and Technology of China, Hefei 230026, China}
\affiliation{Shanghai Branch, CAS Center for Excellence in Quantum Information and Quantum  Physics, University of Science and Technology of China, Shanghai 201315, China}

\date{\today}

\begin{abstract}

Hybrid quantum systems aim at combining the advantages of different physical systems and to produce innovative quantum devices. In particular, the hybrid combination of superconducting circuits and spins in solid-state crystals is a versatile platform to explore many quantum electrodynamics problems. Recently, the remote coupling of nitrogen-vacancy center spins in diamond via a superconducting bus was demonstrated. 
However, a rigorous experimental test of the quantum nature of this hybrid system and in particular entanglement is still missing. We review the theoretical ideas to generate and detect entanglement, and present our own scheme to achieve this.  

\end{abstract}

\maketitle

\section{Introduction}

Quantum information physics has been deeply interdisciplinary from its birth, combining the laws of quantum mechanics with practical fields such as computing \cite{Arute:2019a} and cryptography \cite{Chen:2021a}. Over the past 30 years, quantum information protocols and tasks have been realized in a variety of physical systems, including optical photons, atoms, ions, spins, quantum circuits and many more \cite{Ladd:2010a}. 
Based on the advantages of different systems, the idea of hybrid quantum systems was proposed \cite{Andre:2006q} which aims at coupling different physical systems and use their advantages. Such systems can not only provide new platforms and techniques to study new physics in cavity QED, but are also expected to bring momentum in fields such as quantum computing and quantum enhanced sensing \cite{Xiang:2013a,Kurizki:2015a,Clerk:2020a}. 

One highly potential architecture of hybrid quantum system involves coupling superconducting circuits with spin ensembles, particularly that of negatively charged nitrogen-vacancy centers in diamond (NV$^-$) \cite{Imamoglu:2009a}, which has received wide research interest over the last decade \cite{Schuster:2010a,Kubo:2010a,Amsuss:2011a,Zhu:2011b,Kubo:2011a,Hummer:2012a,Saito:2013a,Grezes:2014a,Putz:2014b,Putz:2017a,Astner:2017a,Angerer:2018a}. In such a system, the strong coupling regime can be achieved as the coupling of an NV ensemble (NVE) with $N$ spins to a single photon can be enhanced by a factor of $\sqrt{N}$ times \cite{Dicke:1954a}. This hybrid system has the combined advantages of fast and efficient tunability, as well as the scalability of superconducting qubits, and the long coherence times of the NV$^-$ spin, thus providing a promising route to realize a hybrid quantum device to read and write quantum information, which is an essential portion of quantum computing, quantum memories, and networking. 

\begin{figure}[b]
\includegraphics[width=0.48\textwidth]{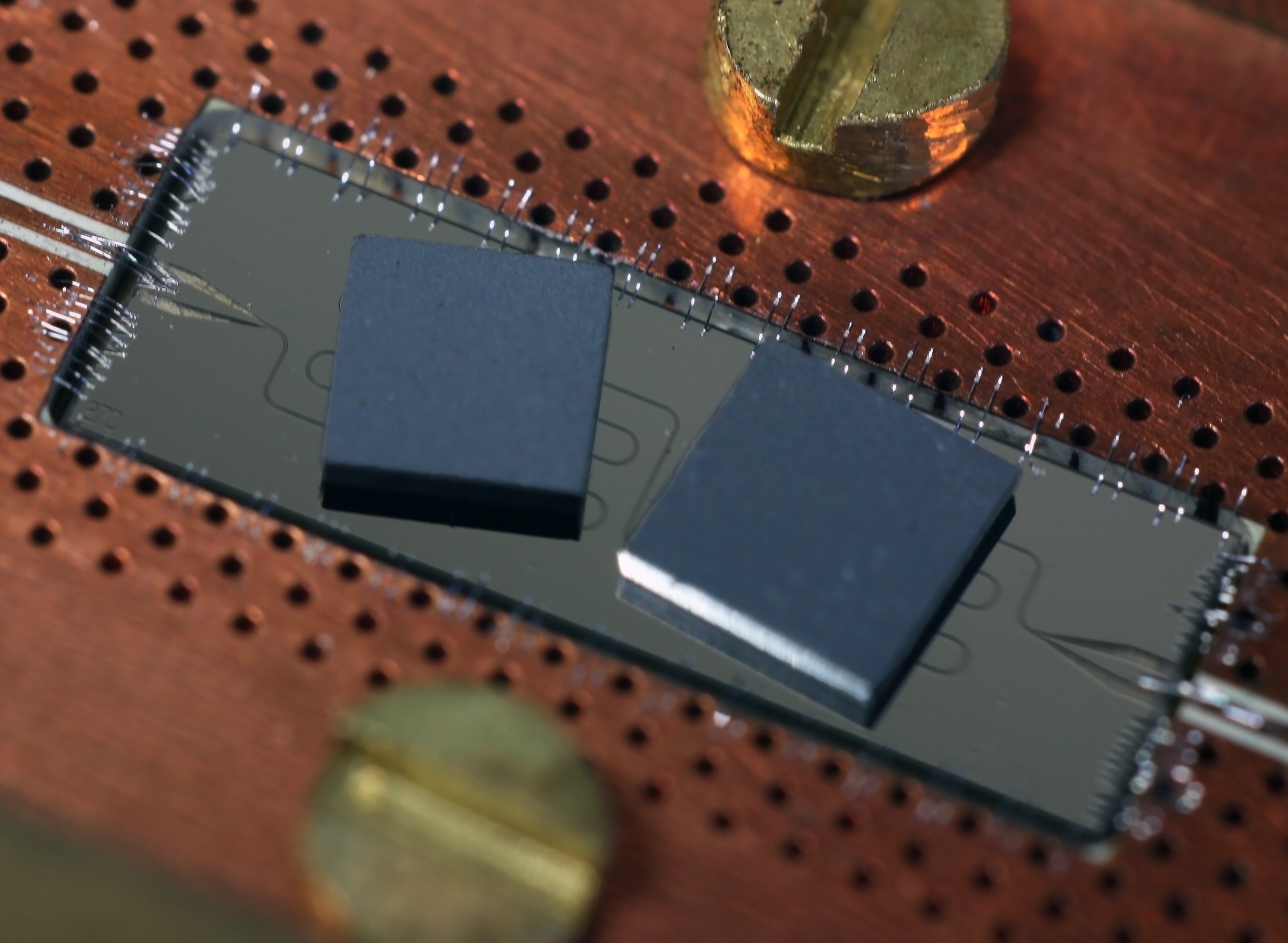}
\caption{Photo of the experimental setup in Ref.~\onlinecite{Astner:2017a}. Two diamonds (black squared crystals) are strongly and coherently coupled via a superconducting niobium chip (4mm $\times$ 12mm). The rotation of the two crystals allows to tune them individually via the magnetic field and study the system coupled as well as uncoupled.
\label{fig:doublediamond}
}
\end{figure}

Meanwhile, entanglement is arguably the hallmark of quantum information\cite{Horodecki:2009a}, as its existence has been proven a necessary condition for achieving quantum advantages for numerous protocols in metrology \cite{Zhang:2015a}, cryptography \cite{Gisin:2002a} and computing \cite{Jozsa:2003a}. Entanglement between single NV$^-$s has been witnessed in the past, such as entanglement over $25$ nm based on the magnetic dipole coupling of two single NV$^-$s at room temperature \cite{Dolde:2013a} and heralding entanglement over $1.3$ km by projecting the photons of spin-photon entangled pairs on to a maximally entangled state \cite{Hensen:2015a}. Although two ensembles of NV$^-$ have been strongly coupled via a superconducting resonator \cite{Astner:2017a} (see Fig.~\ref{fig:doublediamond}), entanglement in such a hybrid quantum system remains to be witnessed. Such correlations could harness spin-based quantum gates \cite{Zu:2014a}, as well as entanglement of quantum memories \cite{Yu:2020a}, which are important for applications in quantum computing and quantum communications. However, how to generate entangled states of remote spin ensembles on a resonator, as well as how to perform independent read out measurements on each NVE is still under investigation. Furthermore, NVEs for hybrid quantum systems contain more than $10^{12}$ spins, which are orders of magnitude larger than that in most previous studies of NV$^-$s \cite{Awschalom:2018a}. A high number of spins leads to inhomogeneous broadening \cite{Wesenberg:2011a}, which results in shorter coherence times and less efficient pulse sequence based measurements, and therefore causes more challenges in witnessing entanglement in such a hybrid quantum system.

The aim to demonstrate entanglement on hybrid quantum systems has triggered active research recently. Theorists have focused on providing schemes for generating entanglement between NVEs \cite{Li:2012a,Yang:2012a,Chen:2012a,Ma:2015a,Song:2015a,Liu:2016a,Song:2016a,Wang:2017a,Maleki:2018a,Su:2018a,Maleki:2018b,Maleki:2019a,Hou:2019a,Teper:2020a}, while experimentalists have developed the necessary state-of-the-art techniques to study this issue \cite{Kubo:2011a,Saito:2013a,Astner:2017a}. In this perspective, we review these works and discuss the possibility of experimentally verifying entanglement between spin ensembles with hybrid quantum systems.

\section{Entanglement properties of the hybrid system}

For the rest of this perspective, we will consider a hybrid system composed of NV$^-$ spin ensembles coupled with superconducting resonators. In particular, we regard the two ensembles as a general bipartite system \cite{Wootters:2001a} and aim at quantifying the entanglement between these remote ensembles.
In order to verify entanglement, first one should inspect the evolution of such a system and theoretically prove that such entanglement correlations exist. This is usually done by calculating the Schrodinger equation to obtain the combined density matrix of the resonator and two ensembles. Then by tracing out the resonator and adopting tools for calculating the degree of entanglement, as well as adjusting parameters such as detuning, collective coupling strength, and frequencies of the two spin ensembles, one can obtain a maximally entangled state that is generated periodically over time in the hybrid system. 

However, while theoretically establishing entanglement in a hybrid system is in principle feasible, an experimental realization is challenging. For instance, the conditions can be arbitrarily set in theoretical proposals - the initial state of the hybrid system is where only one ensemble is excited and both the resonator and the second ensemble are in their ground states. 
Experimentally, this is very difficult to achieve due to the fact that generally the inputs to the resonator are coherent states, which cannot generate entanglement without post-selection. The interaction of the NVEs and the resonator is analogous to the beam splitter in quantum optics \cite{Gerry:2004a}, which requires non-classical sources such as Fock states produce entanglement. 
Meanwhile, to claim entanglement one would need to perform rigorous tests, such as tomography \cite{DAriano:2003a}, entanglement witnesses \cite{Horodecki:1996a,Terhal:2000a}, and so on \cite{Guhne:2009a}, but directly, one can only use the transmission spectrum from the resonator to deduce the system state in hybrid quantum systems. This transmission spectrum is a collective measurement on both ensembles and not a direct measurement on either ensemble.

\section{Theoretical proposals for entanglement generation and detection between NV$^-$ spin ensembles}
\label{sec:TheoreticalProposals}

Theoretically investigating the generation, detection and characterization of entanglement has attracted numerous attention since the beginning of hybrid quantum research, with many schemes having been proposed \cite{Li:2012a,Yang:2012a,Chen:2012a,Ma:2015a,Song:2015a,Liu:2016a,Song:2016a,Wang:2017a,Maleki:2018a,Su:2018a,Maleki:2018b,Maleki:2019a,Hou:2019a,Teper:2020a}. These proposals are mainly focused on how to create quantum states in a hybrid system instead of using coherent states, and how different experimental configurations of the hybrid system can affect the entanglement properties of the two ensembles.

A single spin ensemble couples to either a resonator, as demonstrated in  Ref.~\onlinecite{Kubo:2011a}, or directly to a flux qubit, as shown in Ref.~\onlinecite{Zhu:2011b}. For two ensembles, meanwhile, the system configuration can be much richer and complex, and so here we categorize the proposed schemes into three groups. The first type is where two or more NVEs are coupled to a common coplanar waveguide (CPW) resonator \cite{Liu:2016a,Maleki:2018b,Su:2018a,Maleki:2019a,Hou:2019a}, which is the scenario studied in Ref.~\onlinecite{Astner:2017a}. In those proposals, different input states such as coherent states, squeezed states, or others are used, and the quantum dynamics are analyzed by solving the master equation of the combined system of NVEs and resonator to determine the density matrix. By analyzing the density matrices using various methods of entanglement detection, the presence (and even the amount) of entanglement is determined.

The second type of scheme employs similar structure to the experiments in Ref.~\onlinecite{Zhu:2011b}, where the NVEs are directly coupled to a flux qubit \cite{Song:2015a,Wang:2017a,Maleki:2018a}. Specifically, setups such as two NVEs placed on a flux qubit, or a large flux qubit that interacts with two separate small flux qubits that are each coupled to an NVE have been proposed. Theoretically it was shown that  entanglement can be generated in these systems \cite{Song:2015a,Wang:2017a,Maleki:2018a}, however, challenges might arise experimentally since the flux qubits are currently much smaller in size than the ensembles. 

\begin{figure*}[t]
\includegraphics[width=0.85\textwidth]{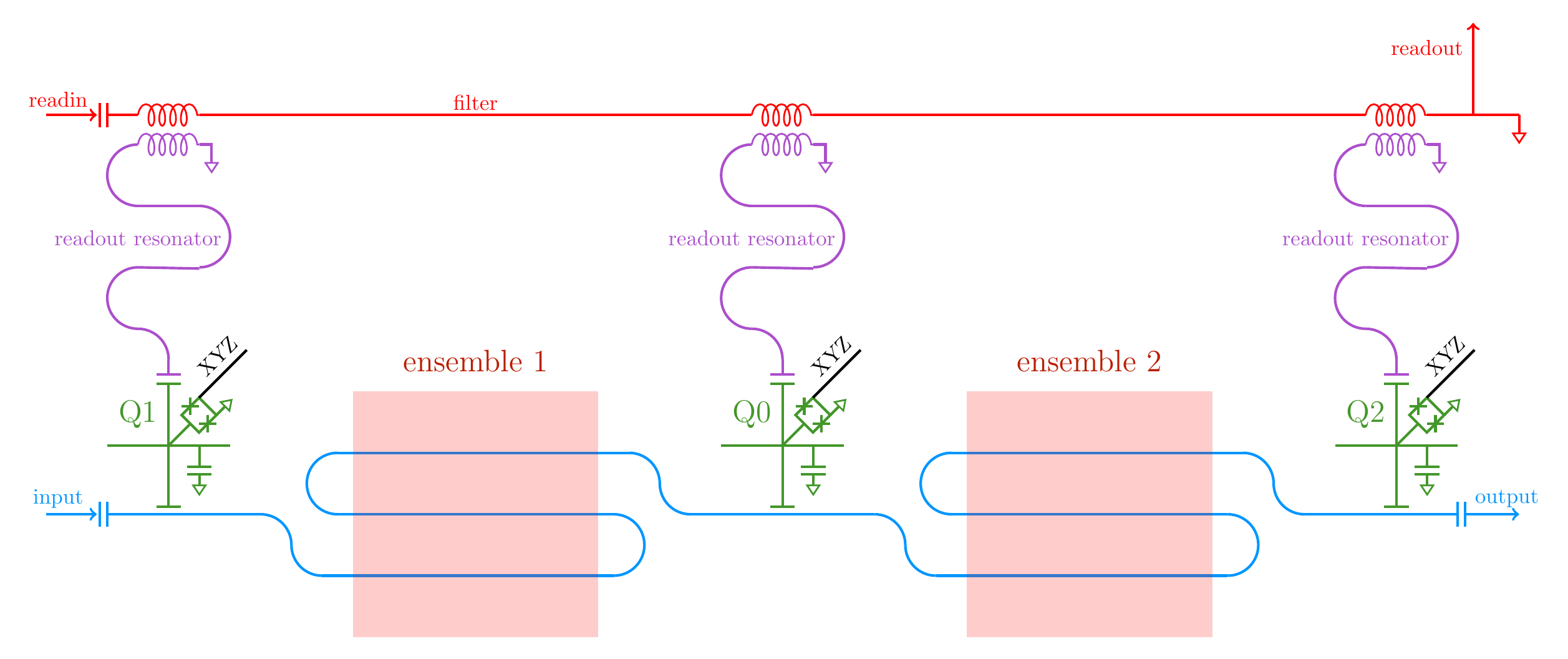}
\caption{Circuit diagram for experimentally verifying entanglement on a hybrid system using a single long resonator. Here the NV$^-$ ensembles are coupling to a long resonator (blue) at the $\lambda/4$ and the $3 \lambda/4$ anti-node of the magnetic field. Three superconducting transmon qubits (green) labelled as Q0, Q1, and Q2, are located at the center and both ends of the resonator coupling to the anti-node of the electric field. Each of these qubits has a separate XYZ control input and a separate readout resonator (purple). All the readout resonators are coupled to a common resonator (red) in order to filter out noise and read out the state of the qubits.
\label{fig:CircuitLongResonator}
}
\end{figure*}

In the third type of scheme, each NVE is coupled to a resonator, while the resonators are connected by an intermediate system, which may be a superconducting quantum interference device (SQUID) or a qubit \cite{Li:2012a,Yang:2012a,Chen:2012a,Ma:2015a,Teper:2020a}. This configuration makes each ensemble directly accessible, and can avoid problems such as frequency crowding which can exist in the first type of configuration, where all components of the hybrid system are integrated onto a single resonator. 
Furthermore, it has been shown that by capacitively coupling such hybrid systems one by one, a chain of entangled ensembles can be generated even in the presence of inhomogeneous broadening \cite{Zou:2014a,Song:2016a}. With this configuration a one dimensional network of entangled quantum memories can be formed \cite{Yang:2011a}, which may bring interesting insight to the Jaynes-Cummings-Hubbard model \cite{Song:2016a}, and provide new platforms to studying phase transitions in driven open quantum systems \cite{Zou:2014a}.

Meanwhile, by adopting different microwave photonic sources, various types of entanglement can be achieved. Using squeezed microwave fields such as Josephson parametric amplifiers (JPA) \cite{Roy:2016a,Hou:2019a,Teper:2020a} and current-biased Josephson-junctions \cite{Yang:2012a}, two-mode squeezed entangled states can be obtained. On the other hand, with flux and transmon qubits \cite{Chen:2012a,Song:2015a,Wang:2017a,Maleki:2018a,Su:2018a} Fock states can be generated, and therewith the existence of highly entangled states between the two ensembles were theoretically shown.

\section{Experimentally verifying entanglement in spin ensemble based hybrid quantum systems}

Although there have been numerous theoretical discussions on entanglement generation and characterization in spin ensemble based hybrid quantum systems, its corresponding experimental realization has not yet been implemented. This gap between theory and experiment can be mainly attributed to challenges in three areas: the initialization or photonic state preparation of the system, the generation of the entangled state, and the detection of its entanglement.  

To create quantum phenomenon such as entanglement between the NVEs, the initial quantum state of the hybrid system should be carefully considered and prepared. In most theoretical proposals, the initial state of the hybrid system is set to only one ensemble excited while both the resonator and the second ensemble are in the ground state. 
However, since most experiments adopt coherent states as the photonic source, it is impossible to obtain such Fock states in practice, which impedes any entanglement generation. Although some theoretical works have suggested adopting JPA and other squeezed microwave field sources\cite{Yang:2012a,Hou:2019a,Teper:2020a}, their compatibility with hybrid quantum systems are yet to be demonstrated. In the short term, it may be simpler and more direct to use a superconducting qubit to generate the initial photonic Fock state \cite{Houck:2007a}, and then transferring the state to the NVE. Here, as another advantage, the qubit being a tunable device can serve as a convenient bridge between the resonator and ensemble, and can be easily be frequency tuned to switch on and off the coupling.

In order to generate such entanglement, the photonic state of the qubit needs to be transferred efficiently to the ensembles.
The simplest way to achieve this is through resonant coupling between the ensemble and the qubit. Here the quantum state can be transferred to the two ensembles by utilizing a $\sqrt{\mathrm{iSWAP}}$ operation, followed by a iSWAP operation \cite{SWAP,Williams:2010a,Mariantoni:2011a}. 
These operations correspond to entanglement creation between the qubit and the first ensemble, and exchanging the quantum state between the qubit and the second ensemble, thus finally entangling the two ensembles. 

Now the ensembles are in principle entangled, however rigorous verification with standard entanglement detection techniques \cite{Friis:2019a} is necessary. In most of the theoretical schemes proposed, the analysis of the hybrid system depends on its density matrix, which indicates that state tomography needs to be performed on the two NVEs. 
However, up to now nobody has experimentally demonstrated the direct state tomography of a NVE and neither has a feasible scheme been proposed. 

\begin{figure*}[t]
\includegraphics[width=0.85\textwidth]{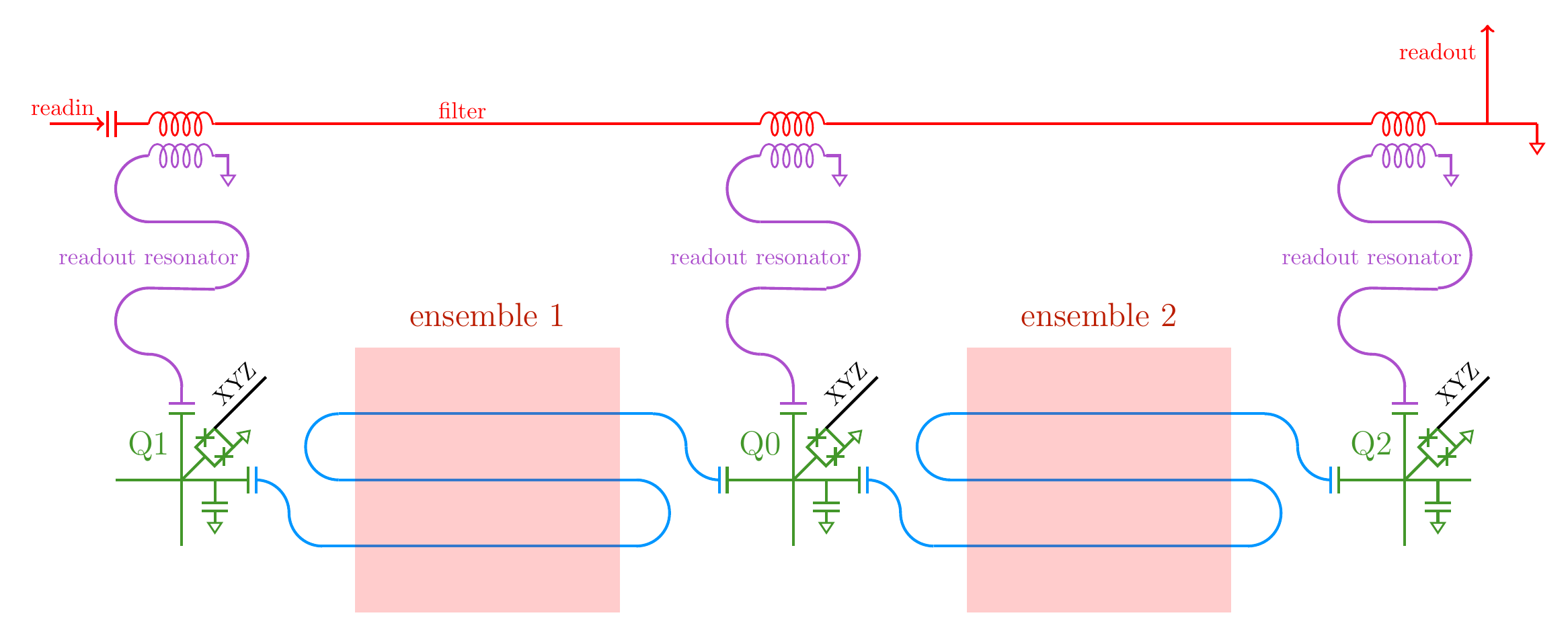}
\caption{Circuit diagram for experimentally verifying entanglement on a hybrid system using a resonator for each spin ensemble. Each NV$^-$ ensemble couples to its separate $\lambda/2$ long resonator (blue). Superconducting transmon qubits (green) are coupled to the end of each resonator. The central qubit Q0 couples to both resonators and allows to transfer quantum information between the two resonators. Exactly like in Fig. \ref{fig:CircuitLongResonator}, each of the qubits has a separate XYZ control input and a separate readout resonator (purple). All the readout resonators are coupled to a common resonator (red) in order to filter out noise and read out the state of the qubits.
\label{fig:CircuitTwoResonator}
}
\end{figure*}

Here, we propose two possible setups for entanglement verification, as shown in Fig.~\ref{fig:CircuitLongResonator} and Fig.~\ref{fig:CircuitTwoResonator}, where we make use of superconducting qubits to retrieve entanglement information of the ensembles. Specifically, we couple ensemble 1 to qubit Q1 and ensemble 2 to qubit Q2, respectively, and then perform entanglement verification on qubits Q1 and Q2 in order to gain information on the quantum state of the two ensembles. 

The difference in the two schemes lies in the setup configuration, where Fig.~\ref{fig:CircuitLongResonator} all ensembles and qubits couple to a long resonator bus, while for the scheme in Fig.~\ref{fig:CircuitTwoResonator} each ensemble couples to an individual resonator. Here the resonators are coupled via a superconducting qubit that is placed between them. Observe that the first scheme (Fig.~\ref{fig:CircuitLongResonator}) is an extended version of the setup in Ref.~\onlinecite{Astner:2017a} with three additional transmon qubits and their readout circuit. This setup has the advantage of a stronger coupling since both ensembles couple to the same resonator. Furthermore, it allows transmission measurements through the long resonator. On the other hand, this may lead to frequency crowding problems, since five systems couple now to a single resonator. The second setup (Fig.~\ref{fig:CircuitTwoResonator}) avoids this issue since each ensemble couples to a different circuit and only three systems couple directly to each resonator. It is worth pointing out that the coupling between the two resonators is indirect and therefore weaker. Furthermore, this coupling can be turned off when necessary with the qubit in between ($\mathrm{Q}_0$) similar to the scheme proposed in reference \cite{Yan:2018a}. In addition, this setup could be extended to more ensembles in a chain like configuration. Note that the first setup follows the first type of theoretical proposals discussed in section \ref{sec:TheoreticalProposals}, while the second setup exemplifies the ideas of the third type of proposals. 

While an experimental verification of entanglement in hybrid system is yet to be realized, the techniques and results shown in current implementation of NVE based hybrid quantum systems are of important reference to discuss the feasibility of our presented schemes. 
For instance, while coherent coupling between a resonator and an NVE is necessary to achieve a hybrid system, the situation is different for entanglement verification, where in order to be entangled the two remote NVEs have to be coherently coupled to each other first.
This was realized in Ref. \onlinecite{Astner:2017a}, where two NVEs were coherently coupled via a resonator, as shown in Fig.~\ref{fig:doublediamond}.
By changing the direction and amplitude of the magnetic field, the two ensembles can be tuned into resonance with the resonator and a large coupling between the three systems was observed. Furthermore, by invoking a large detuning between the resonator and the two ensembles, a dispersive direct coupling between the two ensembles was measured, with only virtually occupying the resonator. This coupling was evidenced by an anti-symmetric state that was observed as a bright state in the dispersive transmission spectrum. 
Notably, when the coupling strengths of two ensembles are set properly, one can derive that the bright state is in principle a maximally entangled state.

The iSWAP operation is crucial in transferring the quantum state between the superconducting qubit and the NVEs, and has been demonstrated in two separated experiments. The first \cite{Kubo:2011a} utilizes a frequency tunable resonator to pass quantum information between a qubit and an ensemble, while the second experiment\cite{Saito:2013a} uses the direct magnetic coupling between an NVE and a superconducting flux qubit. 
A tunable resonator was shown in Ref.~\onlinecite{Kubo:2011a}, where the frequencies of the NVE and transmon qubit are both fixed. Due to flux noise affecting the tunable resonator an adiabatic iSWAP operation was used, where the frequency of the resonator was adjusted adiabatically to transfer quantum states between the resonator and the qubit. Together with state transfer between ensemble and resonator, the authors realized the iSWAP gate in about $500$ ns.
With the alternative route of directly coupling a flux qubit to the spin ensemble, the iSWAP operation is realized through vacuum Rabi oscillations \cite{Saito:2013a}. The advantage of this scheme is that the qubit can couple much stronger to the NVE, which allowed to swap the quantum state between ensemble and qubit in about $30$ ns. 

As a crucial step for the entanglement generation, the $\sqrt{\mathrm{iSWAP}}$ operation was also demonstrated in both experiments. In Ref. \onlinecite{Saito:2013a} the time for performing a  $\sqrt{\mathrm{iSWAP}}$ operation is half of the time needed for a full iSWAP operation, as one expects. On the other hand, in Ref.~\onlinecite{Kubo:2011a} the duration is about the same, since it is mainly dominated by the time needed for the adiabatic pulses.

To readout quantum information from the ensembles, their quantum states are mapped onto qubits Q1 and Q2, where quantum state tomography can be performed to verify entanglement. State tomography of two qubits was first realized in Ref.~\onlinecite{Steffen:2006q}. By measuring the two qubits in combinations of the $X$, $Y$, and $Z$ direction, the full density matrix of the two qubit state can be reconstructed. Here, the measurements can be made by dispersive readout, where the qubit information is extracted from a readout resonator and the results for $|0\rangle$ and $|1\rangle$ can be depicted in two separate clusters. For fast, high-fidelity readout, the single-shot readout using the Purcell filter and parametric amplifier can also be applied. Currently, readout fidelity to $99.2\%$ in $88$ ns can be realized \cite{Walter:2017a}. Using the density matrix, one can examine the entanglement properties of the two ensembles via fidelity and concurrence \cite{Friis:2019a}. Note that the number of measurement settings, or strategies, can be decreased by using entanglement verification methods such as the entanglement witness \cite{Friis:2019a}.

In order to evaluate the feasibility of the experiment, let us consider the following experimental parameters: In a recent experiment \cite{Gong:2021a} a coupling strength of $g_\mathrm{trm}/2\pi=44$ MHz between a transmission line resonator and a transmon qubit was achieved. Furthermore, a coherence time of $T_\mathrm{coh}=13$ $\mu$s was measured, which allows to perform an iSWAP operation with  a fidelity higher than 99\%. The coupling strength between resonator and an NVE was determined to be $g_\mathrm{NVE}/2\pi=9.6$ MHz \cite{Astner:2017a}. The collective polariton mode decay with a rate of $\Gamma/2\pi=1.5$ MHz which estimates the fidelity of the iSWAP to be $85\%$. Additionally, we expect an improvement of these parameters using the electron irradiated samples measured in reference \cite{Astner:2018b} which have a lower inhomogeneous broadening. These experimental parameters suggest that the witnessing of entanglement in this hybrid quantum system is feasible, but only an experiment will confirm it.

\section{Conclusion and outlook}

In this perspective, we have reviewed the developments in witnessing entanglement on hybrid systems. In particular, we investigate the entanglement of two NV$^-$ ensembles via superconducting circuits. We explored the relevant experimental progress towards that goal \cite{Kubo:2011a,Saito:2013a,Astner:2017a} and discussed the related measurements. Furthermore, we reviewed and categorized the theoretical proposals \cite{Li:2012a,Yang:2012a,Chen:2012a,Ma:2015a,Song:2015a,Liu:2016a,Song:2016a,Wang:2017a,Maleki:2018a,Su:2018a,Maleki:2018b,Maleki:2019a,Hou:2019a,Teper:2020a} aiming at entangling ensembles of NV$^-$ spins and presented our own strategy towards that goal. Importantly, although the discussions above are based on the NV$^-$ spin ensemble, the techniques and schemes of entanglement are applicable for any spin ensemble system including magnons \cite{Tabuchi:2014a,Tabuchi:2015a} and molecular spins \cite{Gaita:2019a,Caretta:2021a,Ghirri:2016a}. 

A successful realization of the verification of entanglement in a hybrid system would have a broad impact: Fundamentally, it would demonstrate that a single qubit state could be stored and retrieved from a macroscopic ensemble with of more than a trillion spins. It would further show that two diamonds, two macroscopic object, can be entangled over centimeter distance. This shall enhance the understanding of entanglement manipulation for macroscopic objects \cite{Lee:2011a,Lange:2018a}. 

Finally from the applied point of view, such experiments would demonstrate several key quantum technologies: It shows the controlled storage and retrieval of quantum information in a large ensemble of microscopic spins. It also manifests the controlled creation and manifestation of entanglement in a solid-state system. Such a technology would pave the way for new discoveries and technologies such as quantum simulators, dedicated quantum storage and memories and also quantum transducers.



\begin{acknowledgments}
We thank Yi-Zheng Zhen, Xiaobo Zhu, and Yu-Ao Chen for valuable discussions. 
This work was supported by Anhui Initiative in Quantum Information Technologies, Shanghai Municipal Science and Technology Major Project (Grant No. 2019SHZDZX01), and the Chinese Academy of Sciences.
Y. Mao acknowledges support from the National Natural Science Foundation of China (Grant No. 12104444) and the China Postdoctoral Science Foundation (Grant No. 2021M693093).
W. J. Munro and K. Nemoto acknowledge support from the Japanese JSPS KAKENHI Grant No. 19H00662.

\end{acknowledgments}

\section*{data availability}
The data that support the findings of this study are available from the corresponding author upon reasonable request.

\bibliography{aipsamp}

\begin{thebibliography}{71}
\bibitem [{SWA()}]{SWAP}%
  \BibitemOpen
  \href@noop {} {}\bibinfo {note} {Note that when two energy levels are in
  resonance, any exchange interaction leads to oscillation of the quantum state
  between the two levels. At half of the oscillation period, the states are
  swapped and the operation corresponds to a SWAP gate. At a quarter of the
  oscillation, the operation corresponds to a $\sqrt{\mathrm{SWAP}}$ gate,
  which is an entangling and a universal two-qubit gate.}\BibitemShut {Stop}%
\end{thebibliography}

\end{document}